\documentclass[a4paper, 11pt]{article}
\usepackage[utf8]{inputenc}
\usepackage[T1]{fontenc}
\usepackage[english]{babel}
\usepackage[width=160mm,height=247mm]{geometry}
\usepackage{amsmath, amsfonts, amssymb, bm}
\usepackage[pdftex]{graphicx}
\usepackage{lmodern, indentfirst, cite}
\usepackage[squaren]{SIunits}
\usepackage[textfont={footnotesize},labelfont={color=blue,bf,sf},labelsep=endash]{caption}
\usepackage[colorlinks=true,linkcolor=blue,citecolor=blue,urlcolor=blue]{hyperref}

\usepackage[affil-sl]{authblk}
\title{\textbf{\Large Energy levels of  monolayer -- AA-stacked bilayer graphene quantum dots}}
\author[1]{Abdelhadi Belouad\thanks{\href{mailto:belabdelhadi@gmail.com}{belabdelhadi@gmail.com}}}
\author[1,2]{Abdellatif Kamal\thanks{\href{mailto:abdellatif.kamal@ensam-casa.ma}{
abdellatif.kamal@ensam-casa.ma}}}
\author[1]{El Bouâzzaoui Choubabi\thanks{\href{choubabi@gmail.com}{choubabi@gmail.com}}}
\author[1,3]{Rachid Houça\thanks{\href{mailto:houca.rachid@gmail.com}{houca.rachid@gmail.com}}}
\author[1]{Mohammed El Bouziani\thanks{\href{mailto:elbouziani@yahoo.com}{elbouziani@yahoo.com}}}
\affil[1]{L.P.M.C. Laboratory, Theoretical Physics Group, Faculty of Sciences, Choua\"ib Doukkali University, PO Box 20, 24000 El Jadida, Morocco}
\affil[2]{Department of Mechanical Engineering, National Higher School of Arts and Crafts, Hassan II University, Casablanca, Morocco}
\affil[3]{Equipe de Physique Théorique et Hautes Energies, Faculté des Sciences, université Ibn Zohr, PO Box 8106, Agadir, Maroc}
\date{}
\providecommand{\pacs}[1]{\noindent \textbf{PACS numbers:} #1\\}
\providecommand{\keywords}[1]{\noindent \textbf{Keywords:} #1}
\begin{document}
\begin{titlepage}
	\newgeometry{width=175mm, height=247mm}
    \maketitle
    \thispagestyle{empty}
    \vspace{3cm}
	\begin{abstract}
		This work investigates the electronic properties of the energy spectrum of a hybrid system composed of (i) a circular quantum dot of monolayer graphene surrounded by an infinite sheet of AA-stacked bilayer graphene and (ii) a circular quantum dot of AA-stacked graphene bilayer surrounded by infinite monolayer graphene. We establish analytical findings for the related energy levels and wave functions using the continuum model and the zigzag boundary conditions at the graphene monolayer-bilayer interface. We investigate the effects of perpendicular magnetic, dot radius, and electrical fields on the two types of hybrid system quantum dots. We compare our results to previously published work and explore the potential uses of such a hybrid system quantum dot.
	\end{abstract}
	\vspace{3cm}
	\pacs{81.05.ue, 81.07.Ta, 73.22.Pr}
	\keywords{Graphene, quantum dot, magnetic field, energy levels.}
\end{titlepage}
\restoregeometry
\section{Introduction}
Due to its unique mechanical and electrical characteristics, two-dimensional (2D) carbon crystals such as single layer and bilayer graphene have piqued the curiosity of researcher who wants to utilize them in innovative nanoelectronic devices. The electronic properties of the carriers in the monolayer graphene \cite{Zheng4, Novoselov2, Novoselov5, Zheng5, Gusynin} show an electronic spectrum with no interval and approximately linear near the Fermi energy at two unequal points of the Brillouin zone. The Dirac equation governs graphene charge carriers which are characterized as massless relativistic fermions. In contrast, the spectrum of symmetric graphene bilayer is parabolic around the K points.

\begin{figure}[!htb]\centering
	\includegraphics[scale=0.5]{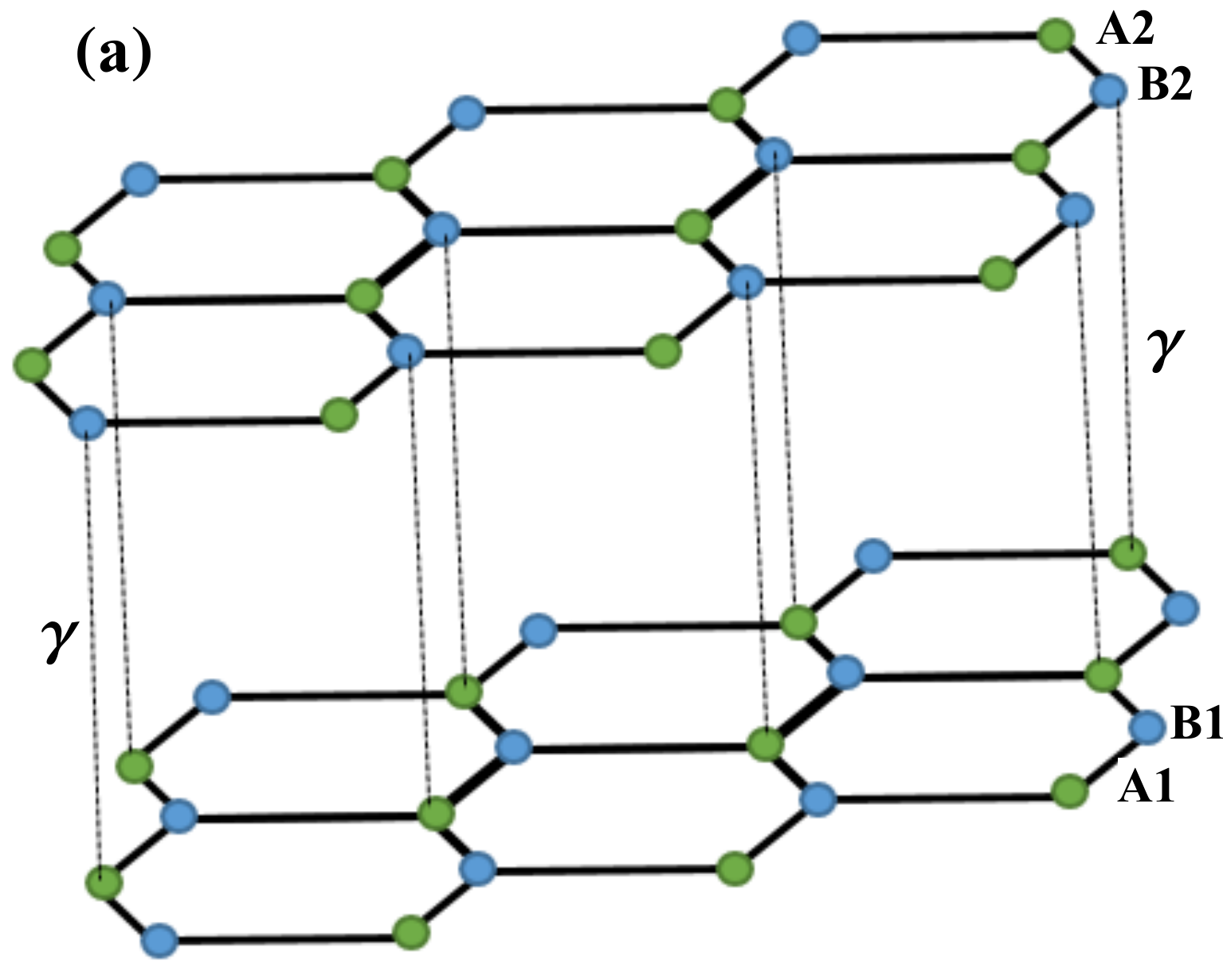}
	\hspace{0.25cm}
	\includegraphics[scale=0.5]{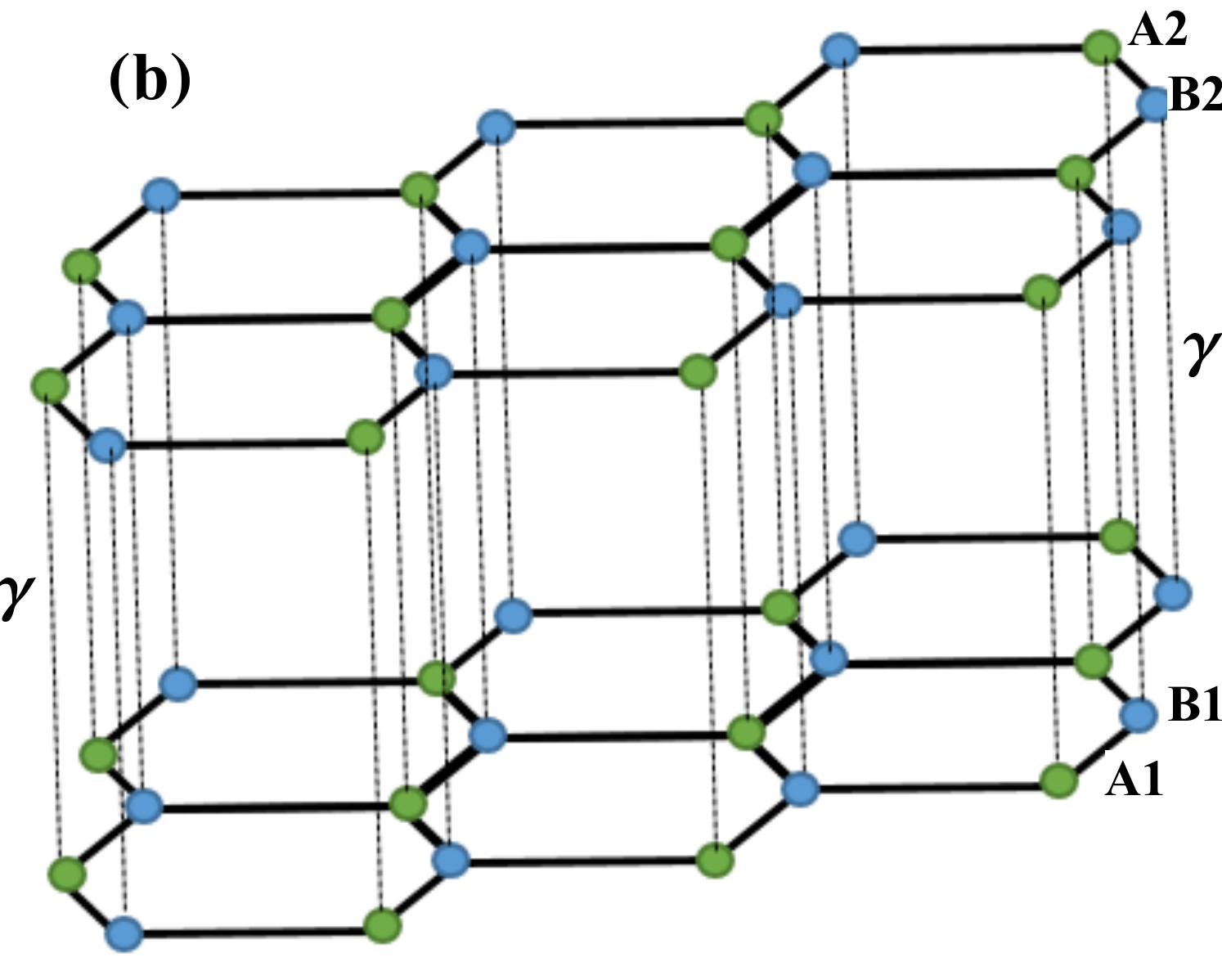}
	\caption{(a) Crystal structure of the AB-stacked bilayer graphene with the parameter $\gamma=400$ meV is the hopping between planes. (b) Crystal structure of the AA-stacked bilayer graphene with the parameter $\gamma=200$ meV is the hopping between planes. The circles denote carbon atoms in the A (green) and B (blue) sublattices in the bottom (1) and top (2) layers. The unit cell of the bilayer graphene consists of four atoms A1, A2, B1, and B2.}
	\label{f1}
\end{figure}

With the richness of interesting properties displayed in graphene, a lot of attention began to shift to the study of bilayer graphene. Two graphene sheets typically take an AB-stacked formation, more commonly know as Bernal stacking. In AB-stacked bilayer, the A atoms in one layer are stacked below the B atoms in the upper layer such that the A atoms in the upper layer sit above the center of the hexagons formed in the lower layer (see Figure \ref{f1}(a)). Like the monolayer, Bernal-stacked bilayer graphene also possesses remarkable properties. The AA-stacked bilayer is characterized by two monolayer sheets stacked directly on top of each other (see Figure \ref{f1}(b)).

The theoretical and experimental work has been done on the AA-stacked bilayer. Recently, however, Lee \textit{et al.} \cite{Lee8} succeeded in growing AA-stacked bilayer on ($111$) diamond. Liu \textit{et al.} have since discovered that bilayer graphene of-ten exhibits AA-stacking but is hard to distinguish from monolayer graphene \cite{Liu9}.

To fabricate and characterize graphene-based nanodevices \cite{Reina9, Soldano10, Choe12}, recent progress has made possible the study of electronic, optical, and transport properties of different graphene-based nanostructures \cite{Novoselov2, Castro Neto9}. In this context, graphene quantum dots (QDs) have been identified as attractive candidates for spin qubits and quantum information storage \cite{Trauzettel7, Silvestrov7}. Because of the Klein tunneling effect, it is well known that lateral confinement of Dirac fermions using an electrostatic gate potential is a difficult operation, which prevents electrical confinement of carriers in graphene \cite{Katsnelson6}. Instead, graphene QDs can be realized by cutting small flakes from a graphene sheet \cite{Schnez9}. It was shown that the energy states of such QDs are strongly dependent on the shape, size, and type of the boundary edges \cite{Bahamon9, Zhang8, Zarenia11, Heiskanen8, Rozhkov11}. On the other hand, it has been demonstrated that for the particular situation of zero-energy modes, bound electron or hole states can be obtained \cite{Calvo11, Bardarson9}.

The paper is organized as follows. In section \ref{sec2}, we consider single layer graphene (SLG)-infinite $AA$-Stacked bilayer graphene (BLG) quantum dot (QD) in section \ref{sec21} absence and in section \ref{sec21} presence of a perpendicular magnetic field. In section \ref{sec3}, we consider $AA$-Stacking bilayer graphene (BLG) quantum dot(QD)-infinite single layer graphene (SLG) quantum dot (QD) in section \ref{sec31} absence and in section \ref{sec32} presence of a perpendicular magnetic field. Section \ref{Results} concerns numerical results for BLG-infinite SLG QDs (SGL-infinite BLG QDs). We conclude the manuscript in section \ref{Conclusion}.
\section{ Energy levels in SLG-infinite AA-Staked BLG QD}\label{sec2}
\begin{figure}[!htb]\centering
	\includegraphics[scale=0.7]{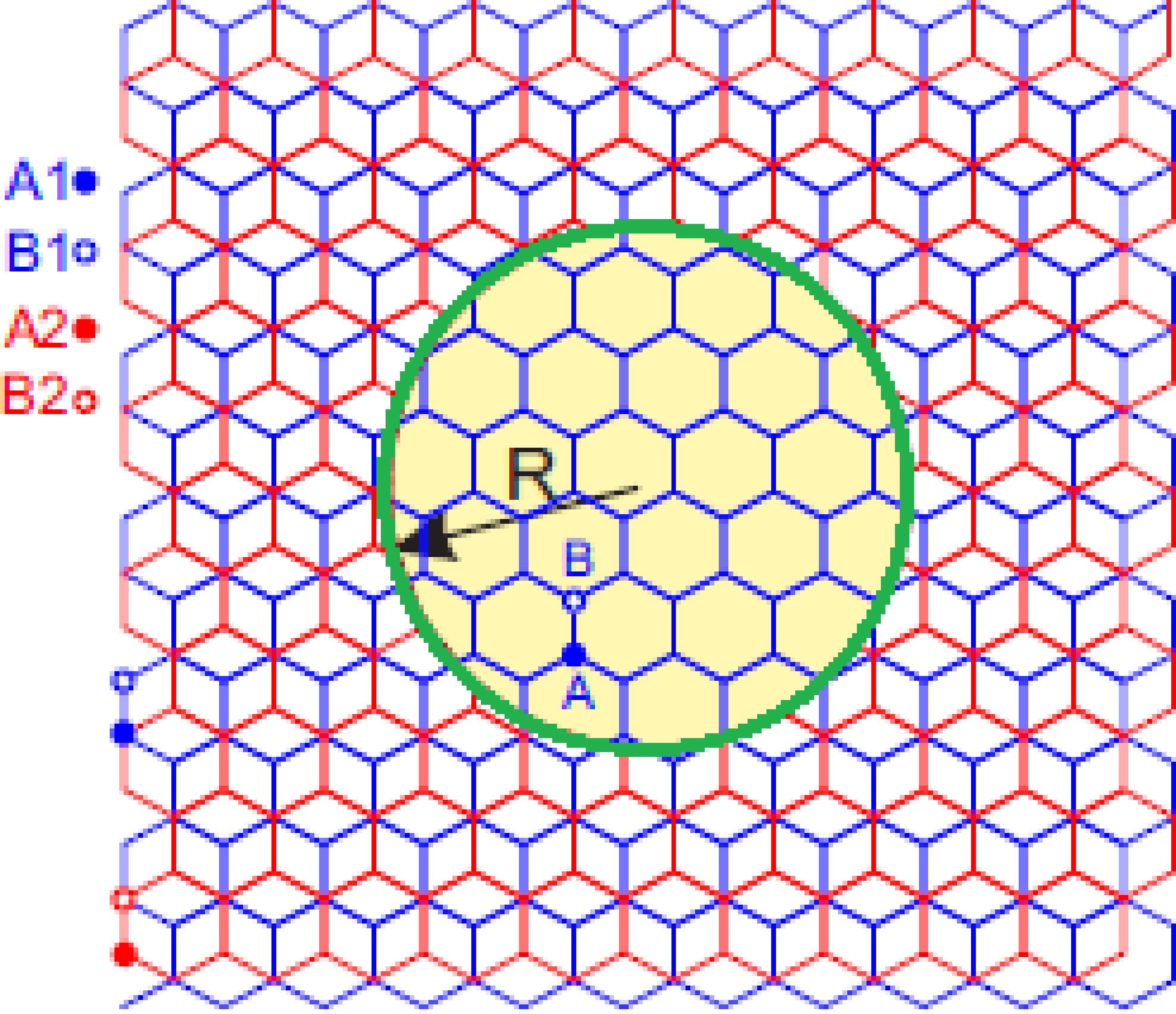}
	\caption{Schematic pictures of the proposed circular SLG-BLG hybrid QDs with radius $R$ (SLG-infinite BLG QD): circular SLG dot surrounded by an infinite BLG.}
	\label{f1S_B}
\end{figure}

In this section, we study the electronic properties of the energy levels of a system consisting of circular SLG QD surrounded by an infinite sheet of AA-staked BLG [see Figure \ref{f1S_B}]. So we assume that one layer of bilayer graphene, containing A1 and B1 sublattices, seamlessly continues to the SLG with $A$ and $B$ sublattices and the second graphene layer is composed of $A2$ and $B2$ sublattices are sharply cut at the boundary $r=R$. We use the corresponding Hamiltonian in both single-layer graphene and bilayer graphene regions and by implementing zigzag boundary conditions to one of the graphene layers at the single-layer graphene-bilayer graphene interface, we calculate the energy levels.

	\subsection{Zero magnetic field}\label{sec21}
		\subsubsection{Single layer graphene}
The dynamics of carriers in a honeycomb lattice of covalent-bond carbon atoms in single layer graphene may be characterized by the Hamiltonian, which is obtained in zero magnetic field by\cite{Recher}
\begin{equation}\label{eq1}
	H=v_F\, \bm{p}\cdot\bm{\sigma}+U_1\,I
\end{equation}
where $v_F = 10^6$ m/s is the Fermi velocity, $\bm{p}=(p_x,p_y)$ is the two-dimensional momentum operator, $\bm{\sigma}=(\sigma_x,\sigma_y)$ are Paulis spin matrices in the basis of the two sublattices of $A$ and $B$ atoms, and $U_1$ is the potential applied to single layer graphene.

We consider that the carriers are confined in a circular area of radius $R$, with zigzag boundary, in polar coordinates, the Hamiltonian \eqref{eq1} reduces to the form
\begin{equation}\label{eq2}
H=\left(%
\begin{array}{cc}
  U_1 & \pi_+ \\
  \pi_- & U_1 \\
\end{array}%
\right)
\end{equation}
where the momentum operators in polar coordinates
\begin{equation}\label{eq3}
	\pi^{\pm} = -i\,\hbar\, v_F\,e^{\pm i\theta}
	\left(
		\frac{\partial}{\partial r}\pm\frac{i}{r}\frac{\partial}{\partial\theta}
	\right),
\end{equation}
$r$ and $\theta $ are polar coordinates. A two-component solution $\Psi$ to the Dirac equation
\begin{equation}\label{eq4}
	H\Psi=E\Psi
\end{equation}
with $E$ being the eigenenergy. The eigenstates of equation \eqref{eq1} are two-component spinors which, in polar coordinates is given by \cite{Zarenia}
\begin{equation}\label{eq5}
	\Psi=e^{im\theta}
		\left(%
			\begin{array}{c}
				\phi_A \\
				i\,e^{-i\theta}\,\phi_B
			\end{array}
		\right)
\end{equation}
with $m=0,\,\pm1,\,\pm2 ...$ being the orbital angular momentum quantum number. The radial components $ \phi_A$ and $\phi_B$ express amplitude probabilities on the two carbon sublattices of graphene, and they satisfy the two coupled differential equations
\begin{eqnarray}\label{eq6}
	&&\left(\frac{\partial}{\partial \rho}+\frac{m}{\rho}\right)\Phi_{A}(\rho)+(\varepsilon-u_1)\,\Phi_{B}(\rho)=0\\
	&&\left(\frac{\partial}{\partial \rho}+\frac{m-1}{\rho}\right)\Phi_{B}(\rho)-(\varepsilon-u_1)\,\Phi_{A}(\rho)=0
\end{eqnarray}
where, in the above equations we used dimensionless units
$$\rho=\frac{r}{R},\qquad\varepsilon=\frac{E}{\hbar\,v_F}R,\qquad u_1=\frac{U_1}{\hbar\,v_F}R.$$

Decoupling the above equations, we arrive at the Bessel differential equation for $\Phi^\tau_{A}(\rho)$
\begin{equation}\label{eq7}
\left[\rho^2 \frac{\partial^2}{\partial \rho^2}+\rho\frac{\partial}{\partial \rho}+ a^2 \rho^2 - m^2\right]\Phi_A^\tau(r)=0.
\end{equation}
where $a=\varepsilon-u_1$, we then find from
\begin{equation}\label{eq8}
    \Phi_{A}(\rho)=C\,J_m(a\rho)
\end{equation}
where  $C$ is the normalization constant. The second component of the wave function can be obtained from \eqref{eq6} as
\begin{equation}\label{9}
    \Phi_{B}(\rho)=- C\, J_{m-1}(a\rho)
\end{equation}
Thus the wave function becomes
\begin{equation}\label{eq5bis}
\Psi=e^{im\theta}
	\left(%
		\begin{array}{c}
			C\, J_{m}(a\rho) \\
			-i\,C\, e^{-i\tau\theta}\,J_{m-1}(a\rho)
		\end{array}
	\right)
\end{equation}
		\subsubsection{$AA$-stacking bilayer graphene}
\begin{figure}[h!]\centering
	\includegraphics[scale=0.5]{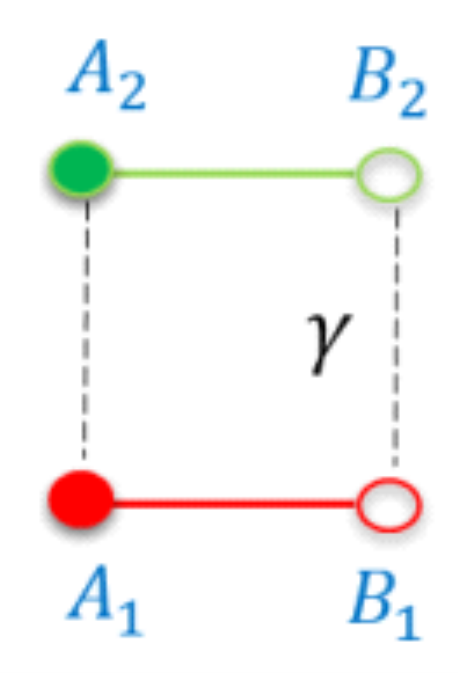}
	\caption{Schematic illustration of the lattice structure of $AA$-stacked bilayer graphene. It consists of two graphene layers. Each carbon atom of the upper layer is located above the corresponding atom of the lower layer and they are separated by an interlayer coupling energy $\gamma$. The unit cell of the AA-stacked bilayer graphene consists of four atoms $A1$, $B1$, $A2$, and $B2$.}
	\label{AA1}
\end{figure}

Let us consider two coupled graphene layers in the $AA$-stacking configuration where the bilayer graphene region can be described in terms of four sublattices, labeled $A1$ and $B1$, for the lower layer $A2$ and $B2$, for the upper layer [see Figure \ref{AA1}]. The $A1$ and $B2$ sites are coupled via a nearest-neighbor interlayer hopping term $\gamma=0.2$ eV. The Bilayer graphene Hamiltonian in the vicinity of the K point is given by \cite{Tabert, Belouad}
\begin{equation}\label{eq10}
	H=H_0+ \dfrac{\Delta U}{2}\sigma_z,
\end{equation}
where
\begin{equation}\label{eq11}
H_0 =
\begin{pmatrix}
U_0 & \pi & 0 & \gamma \\
\pi^\dag & U_0 & \gamma & 0 \\
0 & \gamma &  U_0 & \pi^\dag  \\
\gamma & 0 & \pi & U_0
\end{pmatrix},\qquad U_0=\dfrac{U_1+U_2}{2},\qquad \Delta U=\dfrac{U_1-U_2}{2}
\end{equation}
with $U_1$ and $U_2$ the potentials at the two layers. The operator $\sigma_z$ is defined as
\begin{equation}\label{eq12}
\sigma_z=
\left(%
\begin{array}{cc}
  \textbf{I} & 0 \\
  0 & -\textbf{I} \\
\end{array}%
\right)
\end{equation}
where $\textbf{I}$ is the $2\times2$ identity matrix. The Hamiltonian at the K point is obtained by interchanging $\pi^+$ and $\pi^-$ in \eqref{eq11}. The eigenstates of Hamiltonian \eqref{eq10} are four component spinors \cite{Belouad}

\begin{equation}\label{eq15}
\Psi(r,\theta)=\begin{pmatrix}
					\Phi_A(r)\,e^{im\theta} \\
					i\,\Phi_B(r)\,e^{i(m-1)\theta}\\
					i\,\Phi_{B'}(r)\,e^{i(m-1)\theta}\\
					\Phi_{A'}\,e^{im\theta}
				\end{pmatrix}
\end{equation}
where $m$ is the angular momentum quantum number, which being an integer.

Plugging \eqref{eq15} into the eigenvalue equation $H\Psi(r,\theta)=E\Psi(r,\theta)$, to obtain the following coupled differential equations
\begin{eqnarray}
&& \left(\frac{\partial}{\partial \rho}+\frac{m}{\rho}\right)\Phi_{A}(\rho) + (\alpha-\delta)\,\Phi_{B}(\rho) - \gamma_0\,\Phi_{B'} (\rho)=0 \label{eq161}\\
&& \left(\frac{\partial}{\partial \rho}-\frac{m-1}{\rho}\right)\Phi_{B}(\rho) - (\alpha-\delta)\,\Phi_{A}(\rho) + \gamma_0\,\Phi_{A'}(\rho) \label{eq162}=0\\
&& \left(\frac{\partial}{\partial r}-\frac{m-1}{\rho}\right)\Phi_{B'}(\rho) - (\alpha+\delta)\,\Phi_{A'}(\rho) + \gamma_0\,\Phi_{A}(\rho)=0 \label{eq163}\\
&& \left(\frac{\partial}{\partial \rho}+\frac{m}{\rho}\right)\Phi_{A'}(\rho) + (\alpha+\delta)\,\Phi_{B'}(\rho) - \gamma_0\,\Phi_{B}(\rho)=0 \label{eq164}
\end{eqnarray}
where
$$\gamma_0=\dfrac{R}{\hbar\, v_F}\gamma,\qquad u_{1,2}=\dfrac{U_{1,2}}{\hbar\, v_F} R,\qquad u_0=\dfrac{U_{0}}{\hbar\, v_F} R, \qquad \alpha=\epsilon-u_0,\qquad \delta=\dfrac{u_1-u_2}{2}.$$

Decoupling the system of differential equations \eqref{eq161}-\eqref{eq164} we arrive at the ordinary Bessel differential equation obtain for $\Phi_{A}(\rho)$
\begin{equation}\label{eq17}
	\left[\rho^2 \frac{\partial^2}{\partial\rho^2}+\rho \frac{\partial}{\partial \rho}+ k_{\pm}^2\, \rho^2 -m^2 \right]\Phi_A(r)=0
\end{equation}
where
\begin{equation}
	k_{\pm}=\text{Sign}(\epsilon)\left(\alpha\pm\sqrt{\gamma^2+\delta^2}\right).
\end{equation}
The differential equation \eqref{eq17} is the known modified Bessel equation. Here we choose the modified Bessel function of the second kind $K_m(k_{\pm})$ \cite{Mirzakhani16}, as the appropriate solutions vanishing at $r\longrightarrow\infty$. Thus we have
\begin{equation}\label{eq18}
	\Phi_A(r)=C_1\,K_m(k_+\rho)+C_2\,K_m(k_\rho)
\end{equation}
Using equations \eqref{eq161}-\eqref{eq164}, we obtain the other spinor components
\begin{eqnarray}
	&&\Phi_{A'}(r) = C_1\,\nu_+\,K_m(k_+\rho)+C_2\,\nu_-\,K_m(k_-\rho)  \label{eq191}\\
	&&\Phi_{B}(r) = C_1\,\lambda_+\,K_{m+1}(k_+\rho)+C_2\,\lambda_-\,K_{m+1}(k_-\rho) \label{eq192}\\
	&&\Phi_{B'}(r)=C_1\,\mu_+\,K_{m+1}(k_+\rho)+C_2\,\mu_-\,K_{m+1}(k_-\rho) \label{eq193}
\end{eqnarray}
where
$$\nu_\pm = \frac{\gamma_0^2+(\alpha-\delta)^2+k^2_\pm}{2\, \alpha\,\gamma_0},\qquad
\lambda_\pm=\frac{\alpha+\delta+\gamma_0\,
\nu_\pm(\gamma_0^2+\delta^2-\alpha^2)}{\alpha^2-\gamma_0^2-\delta^2}k_\pm,$$
$$\mu_\pm=\frac{\gamma_0-\nu_\pm(\alpha-\delta)(\alpha^2-\gamma_0^2-\delta^2)}{\alpha^2-\gamma_0^2-\delta^2}k_\pm,$$
and $C_j (j = 1,2)$ are the normalization constants.

Now, we apply zigzag boundary conditions \cite{Mirzakhani16, Koshino10} at the single layer
graphene-$AA$-stacked bilayer gaphene interface. These conditions yield
\begin{eqnarray}
	&& \Psi_A=\Psi_{A1}|_{\rho=1} \label{eq21a}\\
	&& \Psi_B=\Psi^\tau_{B1}|_{\rho=1} \label{eq21b}\\
	&& 0=\Psi_{B2}|_{\rho=1} \label{eq21c}
\end{eqnarray}
the above conditions lead to a system of equations from which we obtain the eigenvalues. For the K point, with the help of the wave functions \eqref{eq5bis}, \eqref{eq15}, and \eqref{eq18}, we arrive at
\begin{equation}\label{eq22}
M_1\left(%
\begin{array}{c}
  C \\
  C_1 \\
 C_2 \\
\end{array}%
\right)=\left(%
\begin{array}{ccc}
  -J_m(a) & K_m(k_+) & K_m(k_-) \\
  J_{m+1}(a) & \lambda_+\,K_{m+1}(k_+) & \lambda_-\,K_{m+1}(k_-) \\
  0 & \mu_+\,K_{m+1}(k_+) & \mu_-\,K_{m+1}(k_-) \\
\end{array}%
\right)\left(%
\begin{array}{c}
  C \\
  C_1 \\
 C_2
\end{array}%
\right).
\end{equation}
The solutions of $\det M_1=0$ are given by
\begin{multline}\label{eq23}
  (\lambda_-\,\mu_+-\lambda_+\,\mu_-)J_m(a)\,K_{m+1}(k_+)\,K_{m+1}(k_-)+\\
 J_{m+1}(a)\Big[\mu_+K_{m+1}(k_+)\,K_{m}(k_-)-\mu_-K_{m}(k_+)\,K_{m+1}(k_-)\Big]=0
\end{multline}
	\subsection{Nonzero magnetic field}\label{sec22}
		\subsubsection{Single layer graphene}
The Dirac Hamiltonian for electron states in graphene, in the presence of a perpendicular magnetic field $B = Bez$ and the potential applied to single layer graphene $U_1$, reads
\begin{equation}\label{eq26}
	H=v_F (\bm{p}+e\,\bm{A})\bm{\sigma}+U_1\,\bm{I}
\end{equation}
where $\bm{A} = (0,B\,r/2, 0)$ is the vector potential in symmetric gauge and $\sigma$ denotes the Pauli matrices, the momentum operators $\pi^{\pm}$ in Hamiltonian \eqref{eq2} are defined as in the presence of a perpendicular magnetic field $B$.
\begin{equation}\label{eq27}
	\pi^{\pm}=-i\,\hbar\, v_F\,e^{\pm i\theta}\left(\frac{\partial} {\partial r}\pm\dfrac{i}{r}\dfrac{\partial}{\partial \theta}\mp i\dfrac{e\,B\,r}{2\,\hbar}\right)
\end{equation}

By acting on the Hamiltonian \eqref{eq26} on $\Psi$, given by the equation \eqref{eq5} we obtain the two coupled differential equations
\begin{eqnarray}
	&&\left(\frac{\partial}{\partial \rho}+\frac{m\tau}{\rho}+\tau \beta \rho\right)\Phi^\tau_{A}(\rho)+(\varepsilon-u_1)\Phi^\tau_{B}(\rho)=0 \label{eq291}\\
	&&\left(\frac{\partial}{\partial \rho}+\frac{m\tau-1}{\rho}-\tau \beta \rho\right)\Phi^\tau_{B}(\rho)-(\varepsilon-u_1)\Phi^\tau_{A}(\rho)=0 \label{eq292}
\end{eqnarray}
where $\beta=\frac{e\,B}{2\,\hbar}R$ is a dimensionless parameter. The coupled equations \eqref{eq291}-\eqref{eq292} are solved analitcally, we arrive at a second-order differential equation for $\Phi_{A}(\rho)$ which depends only on $\rho$,
\begin{equation}\label{eq30}
	\left[\rho^2 \dfrac{\partial^2}{\partial \rho^2}+\rho \dfrac{\partial}{\partial \rho}+ k_{\pm}^2\, \rho^2 -m^2-2\,\beta(m-\tau)+(\varepsilon-u_1)^2\rho^2-\beta^2\rho^4 \right]\Phi_A^\tau(r)=0.
\end{equation}
In order to solve this differential equation, we make the ansatz
\begin{equation}\label{eq31}
	\Phi_{A}(\rho)=\rho^{|m|}\,e^{-\frac{\beta\,\rho^2}{2}}\chi(\rho^{2})
\end{equation}
Substituting \eqref{eq31} into \eqref{eq30} to get
\begin{equation}\label{eqccc}
	\left[x \frac{\partial^2}{\partial x^2}+(b-x)\frac{\partial}{\partial x}-a\right]\chi(x)=0
\end{equation}
where
$$x=\beta r^{2},\qquad b=1+|m|,\qquad a=-\frac{(\varepsilon-u_1)^2 }{4\, \beta}+\frac{m-\tau+|m|+1}{2}.$$
The solution is
\begin{equation}\label{eq34}
	\Phi_A(\rho)=\rho^{|m|}\,e^{-\frac{\beta \rho^2}{2}}C^\tau\tilde{M}(a,b,\beta \rho^2)
\end{equation}
Equation \eqref{eqccc} can be solved to get the following combination
\begin{equation}\label{eq35}
	\Phi_B(\rho)=\dfrac{1}{\varepsilon-u_1}\rho^{|m|}\,e^{-\frac{\beta\rho^2}{2}}\,C^\tau\left[\left(\frac{\tau\, m}{\rho}+\beta\,\tau\, \rho\right)\tilde{M}(a,b,\beta\rho^2)-a\,\tilde{M}(a+1,b+1,\beta \rho^2)\right]
\end{equation}
		\subsubsection{$AA$-stacking bilayer graphene}
In the presence of a perpendicular magnetic field $B$ for $AA$-stacked bilayer graphene and the corresponding wave function are respectively given by equations \eqref{eq10}, \eqref{eq27} and \eqref{eq15}. Solving $H\Psi=E\Psi$  we obtain the following set of coupled differential equations
\begin{eqnarray}
&&\left(\frac{\partial}{\partial \rho}+\frac{m}{\rho}+\beta r\right)\Phi_{A}(\rho) = -(\alpha-\delta)\Phi_{B}(\rho)+ \gamma_0\Phi_{B'} (\rho) \label{eq361}\\
&&\left(\frac{\partial}{\partial \rho}-\frac{m-1}{\rho}-\beta r\right)\Phi_{B}(\rho) = (\alpha-\delta)\Phi_{A}(\rho)- \gamma\Phi_{A'}(\rho) \label{eq362}\\
&&\left(\frac{\partial}{\partial r}-\frac{m-1}{\rho}-\beta r\right)\Phi_{B'}(\rho)=(\alpha+\delta)\Phi_{A'}(\rho)-\gamma_0\Phi_{A}(\rho) \label{eq363}\\
&&\left(\frac{\partial}{\partial \rho}+\frac{m}{\rho}+\beta r\right)\Phi_{A'}(\rho)
=-(\alpha+\delta)\Phi_{B'}(\rho)+\gamma_0\Phi_{B}(\rho). \label{eq364}
\end{eqnarray}
Decoupling the above equations with respect to $\Phi_{A}(\rho)$ we arrive at
\begin{equation}\label{eq37}
	\left[\frac{\partial^2}{\partial \rho^2}+\frac{1}{\rho}\frac{\partial}{\partial \rho}-\left( 2\,\beta\,(m-1)+ \frac{m^2}{\rho^2} + \beta^2\, \rho^2- k_{\pm}^2\right)\right]\Phi_{A}(\rho)= 0
\end{equation}
where $k_{\pm}$ is given in \eqref{eq17} and $\beta=\frac{e\, B\, R}{2\,\hbar}$. In order to solve the differential equations \eqref{eq361}-\eqref{eq362}, we make the following ansatz
\begin{equation}\label{eq38}
	\Phi_{A}(\rho)=\rho^{|m|}\,e^{-\frac{\rho^2 \beta}{2}}\chi(\rho^{2})
\end{equation}
and define a new variable $x=\beta\, r^{2}$. Substituting
\eqref{eq38} into \eqref{eq37} to get
\begin{equation}\label{eq39}
	\left[x \frac{\partial^2}{\partial x^2}+(b-x)\frac{\partial}{\partial x}-n_\pm\right]\chi(x)=0
\end{equation}
which has the solution
\begin{equation}\label{eq41}
	\chi(x)=C_1\, U(n_{+},b,x)+C_2\,U(n_{-},b,x)
\end{equation}
were $ U(n_{\pm},b,x)$ is the regularized confluent hypergeometric function with
\begin{equation}\label{eq40}
 b=1+|m|, \qquad n_{\pm}=-\frac{k_{\pm}^2 }{4\, \beta}+\frac{m+|m|}{2}.
\end{equation}
and $C_j$ ($j=1, 2$) are the normalization coefficients.

In order to find  $\Phi_B$, $\Phi_{B'}$ and $\Phi_{A'}$ we insert the solution for $\Phi_A$ in the differential equations \eqref{eq37}. Using the properties of the regularized confluent hypergeometric function\cite{Mirzakhani16}, this results into
\begin{equation}\label{eq42}
	\Phi_A(\rho)=\rho^{|m|}\,e^{-\frac{\beta \rho^2}{2}}\left[C_1\,U(n_{+},b,\beta \rho^2)+C_2 \,U(n_{-},b,\beta \rho^2)\right]
\end{equation}
\begin{multline}\label{1eq43}
\Phi_B(\rho)=-\zeta\,\rho^{|m|-1}\, e^{-\frac{\beta \rho^2}{2}} \Bigg[C_1\,B_+\left[(m+|m|)\,U(n_{+},b,\beta \rho^2) - 2\,n_{+}\,\beta\,\rho^2\,U(n_{+}+1,b+1,\beta \rho^2)\right]\\
+C_2\,B_-\left[(m+|m|)\,U(n_{-},b,\beta \rho^2) + 2\,n_{-}\,\beta\,\rho^2\, U(n_{-}+1,b+1,\beta \rho^2)\right]\Bigg]
\end{multline}
\begin{multline}\label{2eq43}
\Phi_{B'}(\rho)=-\frac{\zeta}{\gamma_0}\rho^{|m|-1}\,e^{-\frac{\beta\rho^2}{2}}\Bigg[C_1D_+\left[(m+|m|)\,U(n_{+},b,\beta \rho^2) - 2\,n_{+}\,\beta\,\rho^2\, U(n_{+}+1,b+1,\beta \rho^2)\right]\\
+C_2D_-\left[(m+|m|)\,U(n_{-},b,\beta \rho^2) - 2\,n_{-}\,\beta\,\rho^2\,U(n_{-}+1,b+1,\beta \rho^2)\right]\Bigg]
\end{multline}
\begin{equation}\label{3eq43}
	\Phi_{A'}(\rho)=\dfrac{1}{2\,\alpha\,\gamma_0}\rho^{|m|}\,e^{-\frac{\beta\rho^2}{2}}\left[C_1\,C_+\,U(n_{+},b,\beta\rho^2)+C_2\, C_-\,U(n_{-},b,\beta \rho^2)\right]
\end{equation}
where
$$C_\pm=-k_\pm^2+\gamma^2+(\alpha-\delta)^2,\qquad B_\pm=-k_\pm^2+\gamma^2+\delta^2+3\,\alpha^2,$$
$$D_\pm=2\,\alpha\,\gamma^2_0+ (\alpha-\delta)\,C_\pm,\qquad\zeta=\frac{1}{2\,\alpha\,(\alpha^2-\gamma_0^2-\delta^2)}.$$
Applying the boundary conditions \eqref{eq21a}-\eqref{eq21c} at the interface $\rho=1$, we arrive at
\begin{equation}\label{eq44}
M_2\left(%
\begin{array}{c}
  C \\
 C_1 \\
  C_2 \\
\end{array}%
\right)=0
\end{equation}
where
\begin{equation}\label{eq45}
M_2=\left(%
\begin{array}{ccc}
  m_{11} & m_{12} & m_{13} \\
  m_{21} & m_{22} & m_{23} \\
  m_{31} & m_{32} & m_{33} \\
\end{array}%
\right)
\end{equation}
with the matrix elements
\begin{eqnarray*}
&& m_{11}=-\tilde{M}(a,b,\beta),\\
&& m_{31}=0,\\
&& m_{12}=U(n_+,b,\beta),\\
&& m_{13}=U(n_-,b,\beta),\\
&&m_{21}=\frac{1}{\varepsilon-u_1}\left[a\,\tilde{M}(a+1,b+1,\beta)-\tau\,(m+\beta)\,\tilde{M}(a,b,\beta)\right],\\
&& m_{22}=\zeta\, B_+\left[(m+|m|)\,U(n_{+},b,\beta) - 2\,n_{+}\,\beta\, U(n_{+}+1,b+1,\beta)\right],\\
&& m_{23}=\zeta\, B_-\left[(m+|m|)\,U(n_{-},b,\beta) - 2\,n_{-}\,\beta\, U(n_{-}+1,b+1,\beta)\right],\\
&& m_{32}=\zeta\, D_+\left[(m+|m|)\,U(n_{+},b,\beta)-2\,n_{+}\,\beta\, U(n_{+}+1,b+1,\beta)\right],\\
&& m_{33}=\zeta\, D_-\left[(m+|m|)\,U(n_{-},b,\beta) - 2\,n_{-}\,\beta\, U(n_{-}+1,b+1,\beta)\right].
\end{eqnarray*}
The energy levels are obtained from the condition $\det|M_2|=0$, which is written as
\begin{equation}\label{eq46}
	\dfrac{m_{13}\,m_{32}  - m_{12}\,m_{33}}{m_{11}} =  \dfrac{m_{23}\,m_{32}-m_{22}\,m_{33}}{m_{21}}
\end{equation}
\section{Energy levels in $AA$-Staked BLG-infinite SLG QD}\label{sec3}
	\subsection{Zero magnetic field}\label{sec31}
\begin{figure}[!htb]\centering
	\includegraphics[scale=0.7]{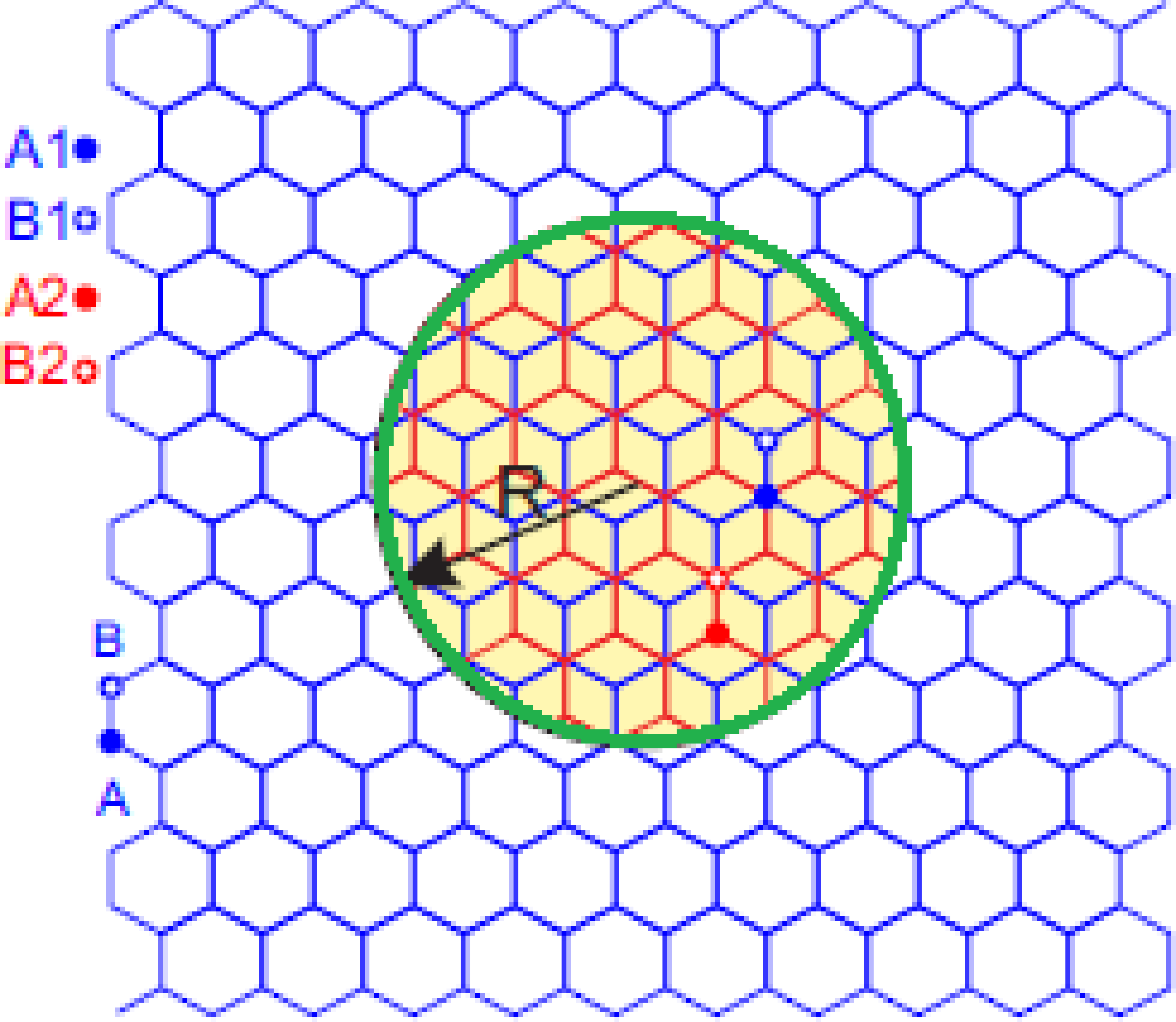}
	\caption{Schematic pictures of the proposed circular BLG-SLG hybrid QDs with radius R. BLG-infinite SLG QD: circular SLG dot surrounded by an infinite BLG.}
	\label{f1B_S}
\end{figure}

In this section, we will consider the inverse of the previous QD system, in which a BLG QD is surrounded by an infinite SLG [see Figure \ref{f1B_S}]. This system is considered as an infinite BLG sheet in which a circle of radius $R$ from its upper layer is left and the other part is removed. The Hamiltonian is calculated for both portions of the system in the same way as in the previous section. Then appropriate wave functions are chosen to meet the extreme requirements when $r\rightarrow0$ for bilayer dot and $r\rightarrow\infty$ for SLG. The solutions for areas of the BLG QD and infinite SLG, respectively, are the modified Bessel function of the first kind $I_m$ and the Bessel function of the second type $Y_m$. Even in the presence of bias, there is no unique linear combination of the real and the imaginary part of $I_m(k_{\pm})$  and $Y_m(k_{\pm})$ from which unique discrete energies may be derived as previously mentioned (See section \ref{sec21}).
	\subsection{Non zero magnetic field}\label{sec32}
In the presence of a magnetic field, the calculations are similar to those presented in section \ref{sec22}. The Hamiltonian is solved for both parts of the system. Then we choose the appropriate wave functions to satisfy the extreme conditions when $ r\rightarrow 0 $ for the bilayer point and $r\rightarrow\infty $ for SLG. Here, the hypergeometric function $M$ and $U$ are respectively the solutions for BLG QD and infinite SLG regions \cite{Mirzakhani16}.

So the elements of the matrix $M^K_{zz}$ becomes

\begin{eqnarray*}
&& m_{11}=-U(a,b,\beta),\\
&& m_{31}=0,\\
&& m_{12}=\tilde{M}(n_+,b,\beta),\\
&& m_{13}=\tilde{M}(n_-,b,\beta),
\end{eqnarray*}
\begin{eqnarray*}
&& m_{21}=\frac{1}{\varepsilon-u_1}\left[a\,U(a+1,b+1,\beta)-\tau(m+\beta)\,U(a,b,\beta)\right],\\
&& m_{22}=\frac{B_+}{2\,\alpha\,(\alpha^2-\gamma_0^2-\delta^2)}\left[(m+|m|)\tilde{M}(n_{+},b,\beta)-2\,n_{+}\,\beta\, \tilde{M}(n_{+}+1,b+1,\beta)\right],\\
&& m_{23}=\frac{B_-}{2\alpha(\alpha^2-\gamma_0^2-\delta^2)}\left[(m+|m|)\,\tilde{M}(n_{-},b,\beta) - 2\,n_{-}\beta\, \tilde{M}(n_{-}+1,b+1,\beta)\right],\\
&& m_{32}=\frac{D_+}{2\,\alpha\,(\alpha^2-\gamma_0^2-\delta^2)}\left[(m+|m|)\,\tilde{M}(n_{+},b,\beta)-2\,n_{+}\,\beta \tilde{M}(n_{+}+1,b+1,\beta)\right],\\
&& m_{33}=\frac{D_-}{2\,\alpha(\alpha^2-\gamma_0^2-\delta^2)}\left[(m+|m|)\,\tilde{M}(n_{-},b,\beta) - 2\,n_{-}\,\beta\, \tilde{M}(n_{-}+1,b+1,\beta)\right].
\end{eqnarray*}
For $K$ valley, the energy levels are obtained from the condition det$|M^K|=0$. We arrive at
\begin{equation}\label{eq46}
	\dfrac{m_{13}\,m_{32}  - m_{12}\,m_{33}}{m_{11}} =  \dfrac{m_{23}\,m_{32}-m_{22}\,m_{33}}{m_{21}}
\end{equation}
\section{Numerical results}\label{Results}
\begin{figure}[!htb]
  \includegraphics[scale=0.5]{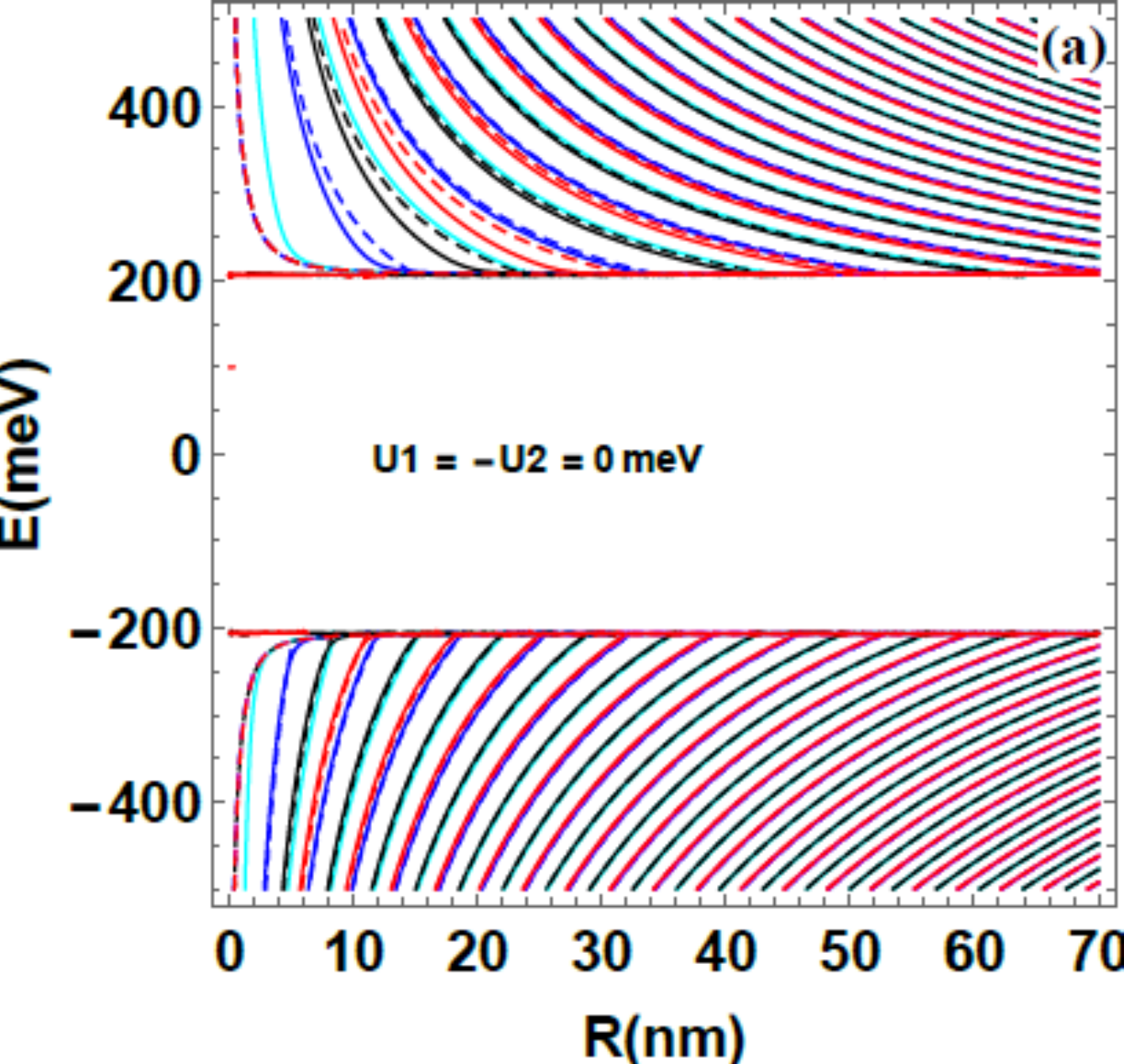}
  \hspace{0.2cm}
  \includegraphics[scale=0.5]{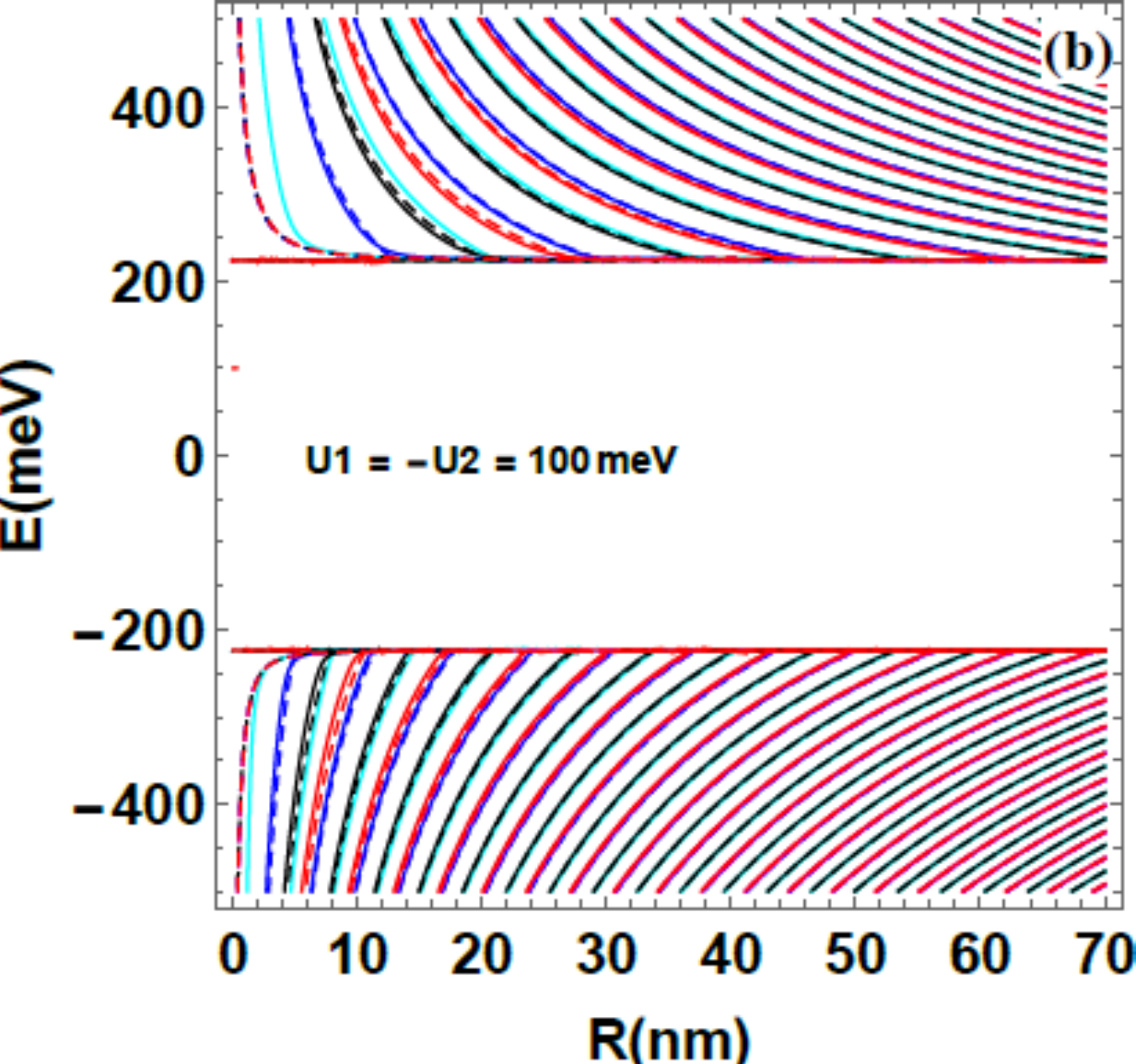}
  \caption{Energy levels of single layer graphene-infinite AA stacking bilayer graphene quantum dot as a function of dot radius $R$ with $B=0$ T, for the angular momenta $m =0$ (cyan solid), $m=1$ (blue dashed), $m=-1$ (blue solid), $m=2$ (black dashed), $m=-2$ (black solid), $m=3$ (red dashed), and $m = -3$ (red solid), with (a): $U_1=U_2= 0$ meV and (b): $U_1=-U_2=100$ meV.}
  \label{f03}
\end{figure}

Figure \ref{f03} show the energy spectrum as a function of the dot radius $R$, at the $K$ valley with $B=0$ T for $(a)$ $U_1=U_2=0$ meV and $(b)$ $U_1=-U_2=100$ meV. These results are for the angular momentum values $m =0$ (cyan), $m=1$ (blue dashed), $m=-1$ (blue), $m=2$ (black dashed), $m=-2$ (black solid), $m=3$ (red dashed), and $m = -3$ (red solid). We show an energy gap appears between bands of conduction and valence depends on $\Delta U$, i.e.  when $\Delta U\rightarrow 0$, $E_g\rightarrow 2\gamma$ as shown in Figure \ref{f03}(a), and when $\Delta U\rightarrow 200$, $E_g\rightarrow 2\gamma=400$ meV  as shown in Figure \ref{f03}(b).
This gap of energy is given by
$$E_g=2\left( \dfrac{\Delta U}{\gamma}+\sqrt{\gamma^2+{\Delta U}^2}\right).$$
the energy level corresponds to the band of valance converges to
$$E_g=-\left(\frac{\Delta U}{\gamma}+\sqrt{\gamma^2+{\Delta U}^2}\right)$$
when the radius the quantum dot increases, on the other hand, the energy level corresponds to the conduction band converges to
$$E_g=\frac{\Delta U}{\gamma}+\sqrt{\gamma^2+{\Delta U}^2}$$
when the radius the quantum dot increases. On the one hand, these results are similar to those obtained for a single layer (SLG)-AB-stacked bilayer graphene (BLG) quantum dots (SLG-Infinite BLG quantum dots)\cite{Mirzakhani16}. On the other hand, they are different from SLG \cite{Gruji} and BLG QD \cite{Pereira09}, this difference is due to the zigzag boundary condition applied to the SLG-BLG interface, which removes the layer symmetry in BLG. Our results show that energy levels verifies $E_k(m)\neq E_k(-m)$.

\begin{figure}[h!]
  \includegraphics[scale=0.5]{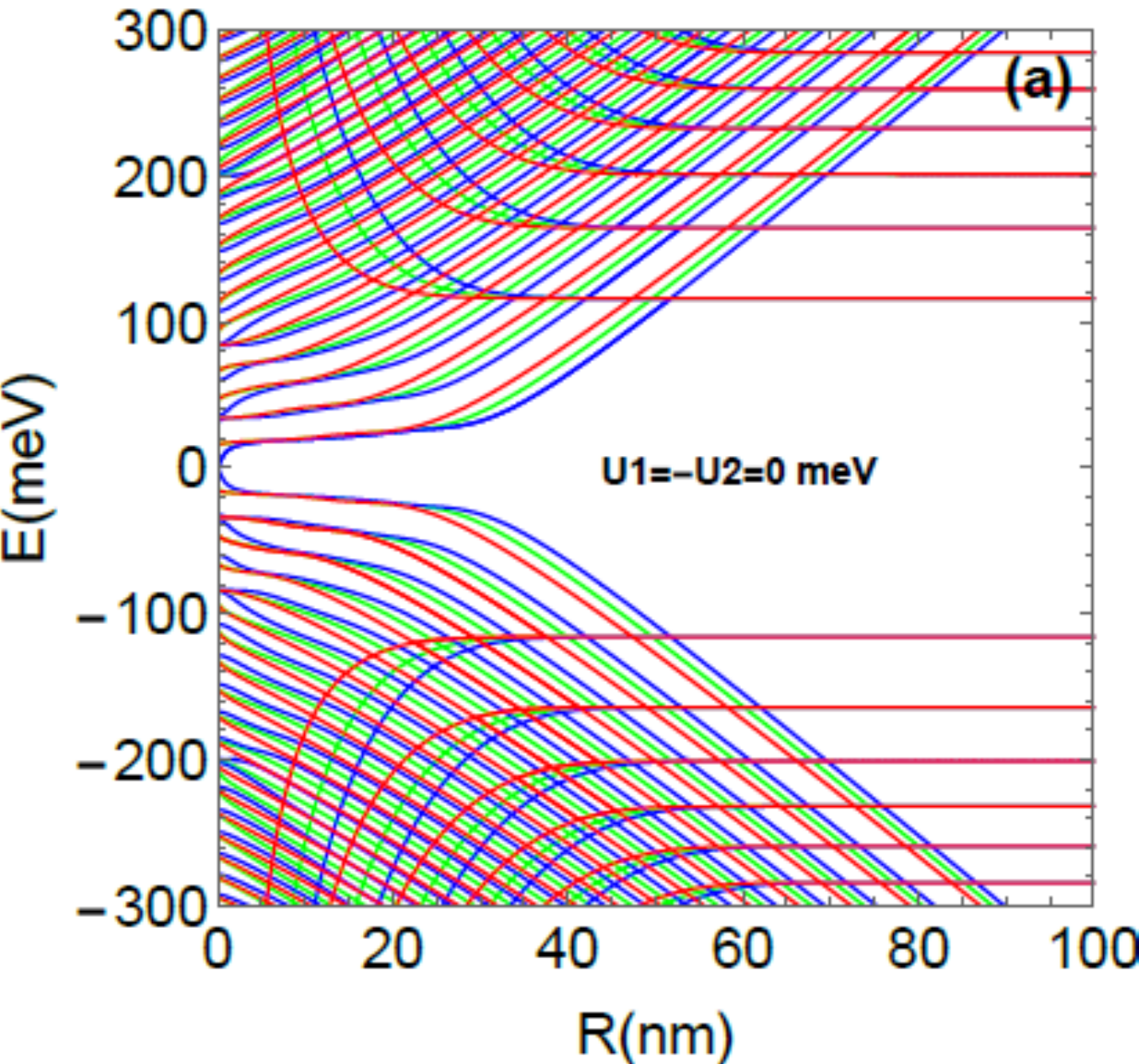}
  \hspace{0.2cm}
  \includegraphics[scale=0.5]{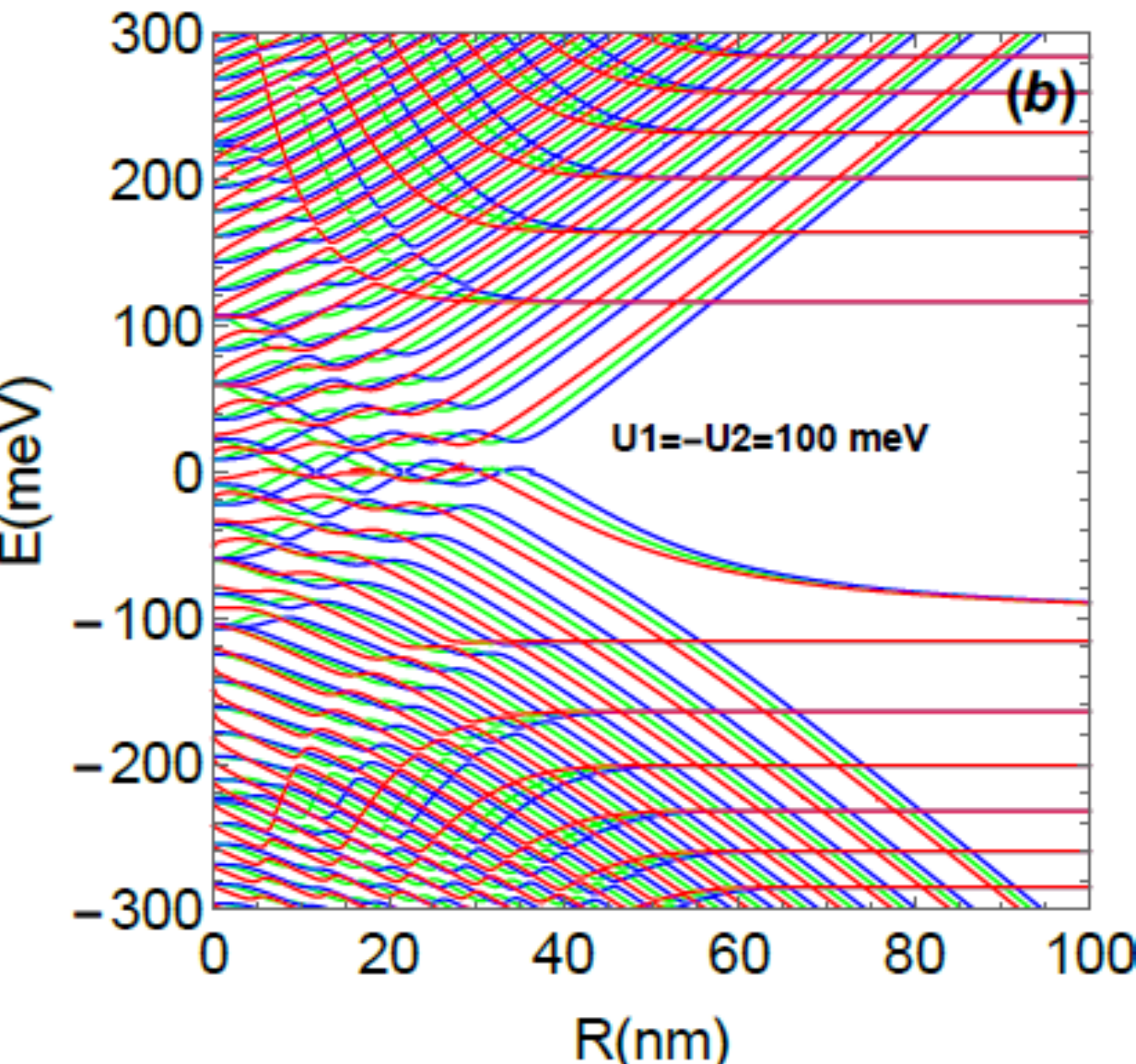}
  \caption{Energy levels of single layer graphene-infinite AA-stacking bilayer graphene quantum dot as a function of dot radius $R$ with $B=10$ T, for the angular momenta $m =0$ (blue solid), $m=1$ (green solid) and $m=-1$ (red), with (a): $U_1=U_2= 0$ meV, (b): $U_1=-U_2=100$ meV at the valleys $K$.}
  \label{f04}
\end{figure}

Figure \ref{f04} shows the energy levels as a function of the dot radius $R$ for three angular momenta, $m=-1$ (blue), $m=0$ (red), $m=1$ (green)  for non zero magnetic field $B=10$ T. In the case $U_1=U_2=0$, the spectrum of Figure \ref{f04}(a) shows a band with gapless between the valence and conduction bands, which opens when the radius of the quantum dot increases, except for the case, $R\approx 0$ and  $E_k($m=0$)= 0$. In contrast, in the case of the non zero potential ($U_1=-U_2=100$ meV), the results show a strong dependence on the quantum dot radius, with the appearance of several crossings when $0\,\text{nm}\leq R\leq 15$ nm, then the appearance of a gap when  $R>15$ nm, as shown more clearly in Figure \ref{f04}(b), furthermore  we see that the energy spectrum shows two sets of levels, a set for $R\rightarrow 0$, the energy levels correspond to the LLs of bilayer graphene given by
\begin{equation}
	\epsilon=u_0\pm\left(\sqrt{\gamma_0^2+\delta^2}\pm\sqrt{2\,\beta\, n+m+\mid m\mid}\right)
\end{equation}
where $n=0,\,1,\,2,\,3,\ldots$. Increasing $R$, the $2^{th}$ set split for different angular momenta and approach the LLs of SLG \cite{S. Schnez}, this behavior is qualitatively similar to that found in  SLG-BLG  \cite{Mirzakhani16}.
\begin{figure}[h!]
	\includegraphics[scale=0.5]{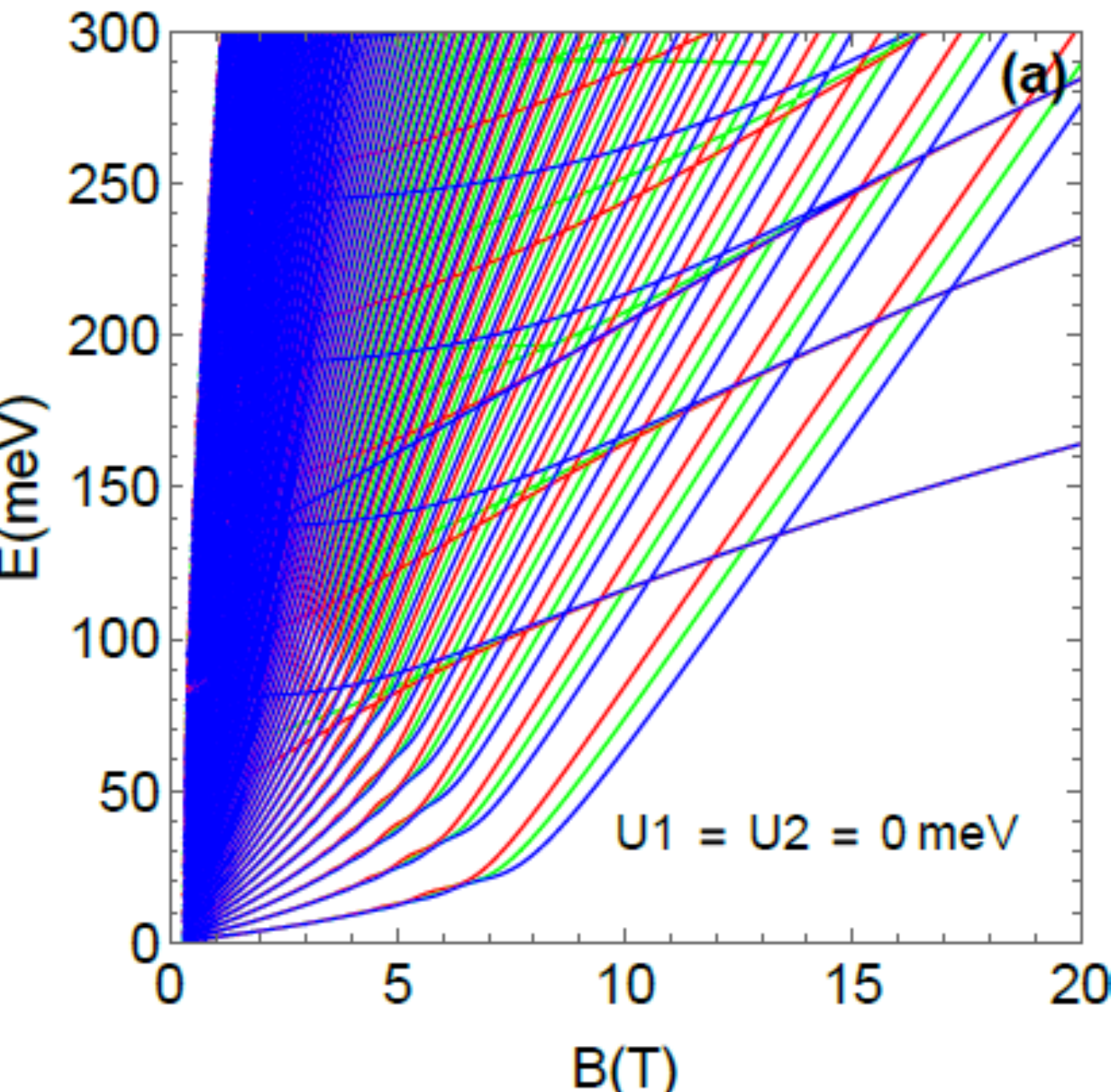}
	\hspace{0.2cm}
	\includegraphics[scale=0.5]{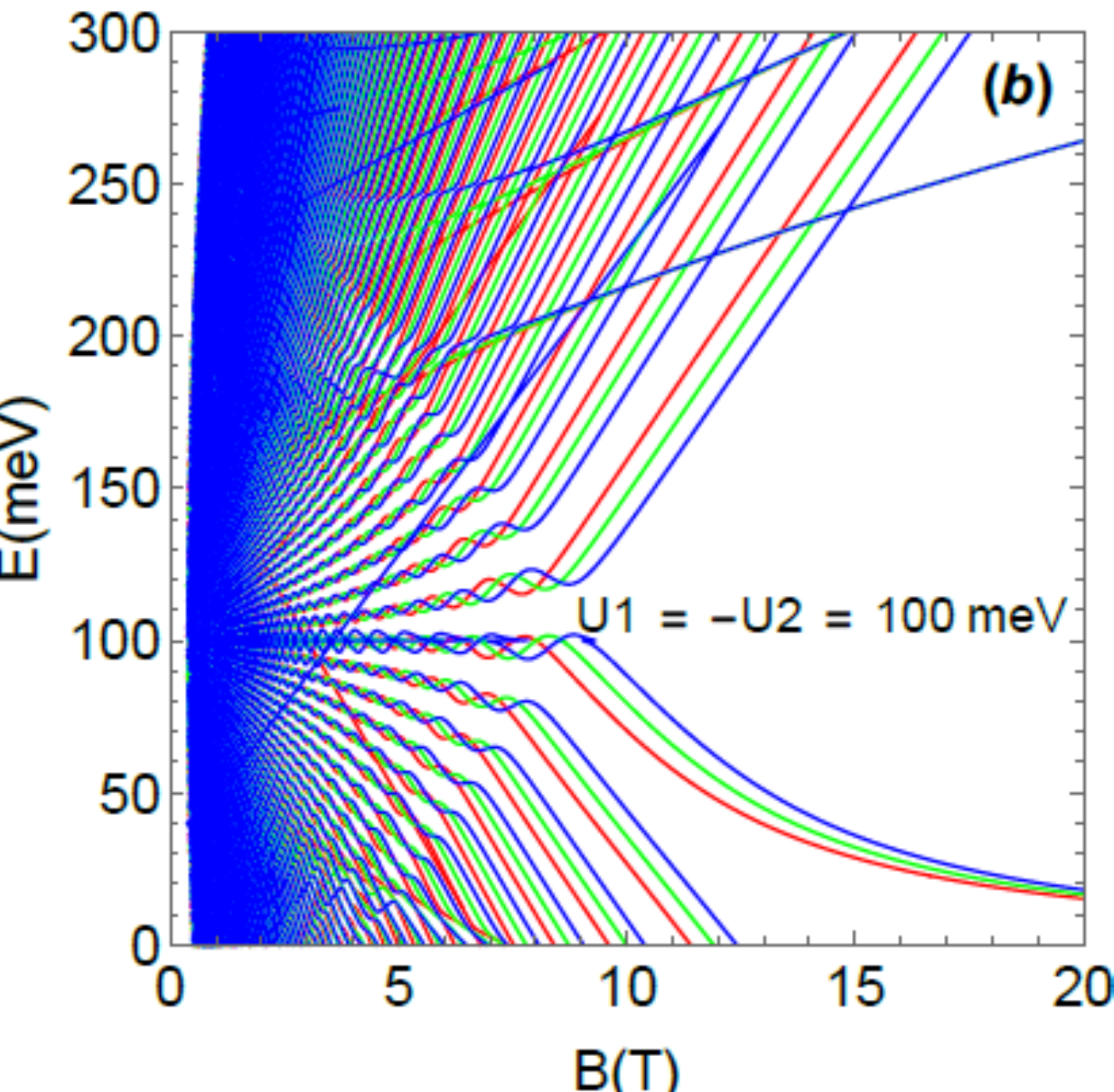}
	\caption{Energy levels of single layer graphene-infinite $AA$-stacking bilayer graphene quantum dot as a function of magnetic field $B$ with $R = 40$\,nm, for the angular momenta $m =0$ (blue), $m=1$ (Red), $m=-1$ (Green), (a): $U_1=U_2= 0$ meV, (b): $U_1=-U_2=100$ meV at the valleys $K$.}
	\label{f05}
\end{figure}

In Figure \ref{f05}, we show the energy levels of single layer-infinite $AA$ stacking bilayer graphene quantum dot as a function of the magnetic field $B$ for (a): $U_1=U_2=0$ meV and (b): $U_1=-U_2=100$ meV, the energy levels are plotted for angular momentum labeled $m=-1$ (blue), $m=0$ (green), and $m=1$ (red) with $R=40$ nm. The energy spectra show two types of energy levels, for small magnetic fields ($B\longrightarrow 0$), this case presents the degeneracy of lower energy states and the spectrum becomes strongly dependent on $m$, ie $E(m)=E(-m)$. However, as the magnetic field increases, the magnetic confinement becomes important, as evidenced by the lifting of the degeneracy of the states, and the energy levels approach the LLs of SLG \cite{Mirzakhani16}. Note that when the application of potential $U_1=-U_2 =100$ meV gives a vertical energy level shift of the order of $100$ meV (Figure \ref{f05}(b)) \cite{youness17}.
\begin{figure}[h!]
	\includegraphics[scale=0.5]{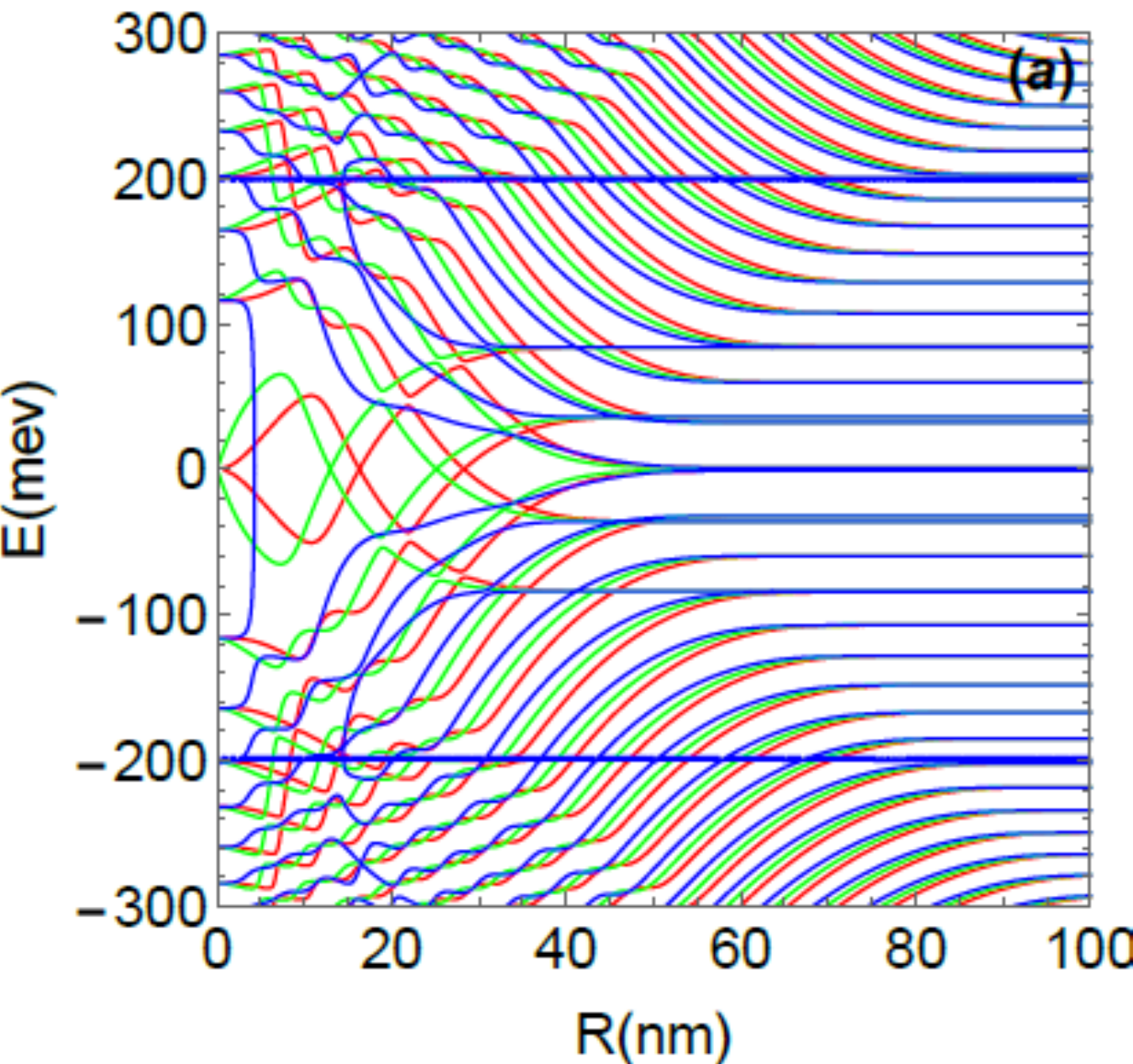}
	\hspace{0.2cm}
	\includegraphics[scale=0.5]{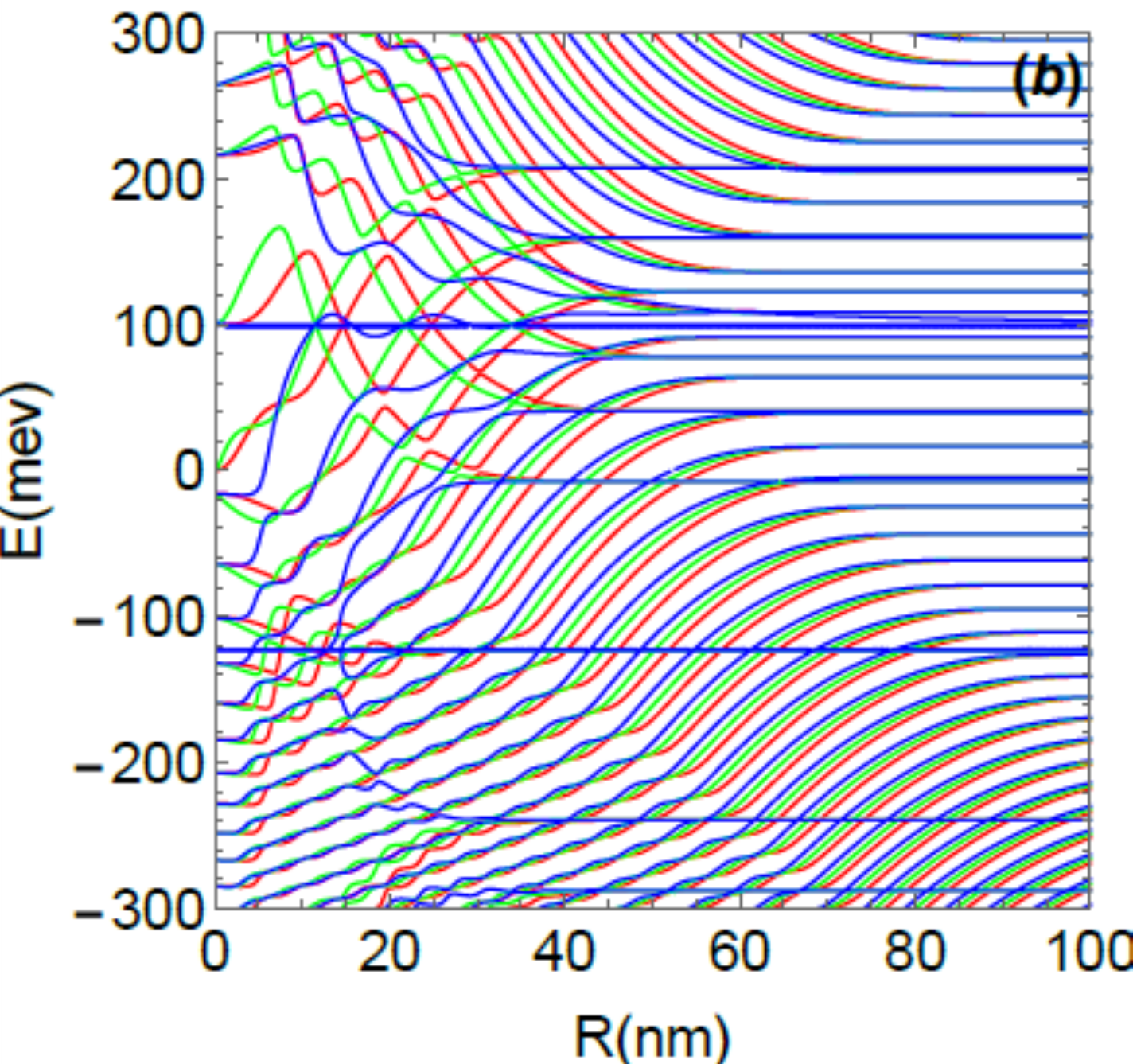}
	\caption{Energy levels of $AA$-stacking bilayer graphene-infinite single layer graphene quantum dot as a function of $R$ with $B = 10$ T, for the angular momenta $m =0$ (blue), $m=1$ (Red), $m=-1$ (Green), (a): $U_1=U_2= 0$ meV, (b) $U_1=-U_2=100$ meV at the valleys $K$.}
	\label{f06}
\end{figure}

Figure \ref{f06} shows the energy levels for $AA$-stacking bilayer graphene (BLG)-infinite single layer graphene quantum dot  (SLG QD) as a function of radius R for the two cases: $U_1 = U_2 = 0$ meV (Figure \ref{f06}(a) and $U_1=-U_2=100$ meV (Figure \ref{f06}(b), the spectrum is plotted for $m=-1$ (blue), $m=0$ (green) and $m=1$ (red), magnetic field $B=10$ T. When $R\rightarrow 0$, the spectrum is in agreement with LLs of SLG with degenerate states for all angular momentum. With increasing point radius, the degeneracy of the levels for different $m$ is lifted and eventually merges into the LLs of BLG, The way the LLs of SLG and BLG are connected is similar to those discussed for the AB-stacked BLG-SLG QD \cite{Mirzakhani16}. The spectrum as a function of $R$ for the biased case with $U_1=-U_2=100$ eV, as shown in Figure \ref{f06}(b) shows the same behavior as the unbiased case $U_1=U_2=0$ eV with a vertical shift of the energy spectrum.
\begin{figure}[!htb]
	\includegraphics[scale=0.5]{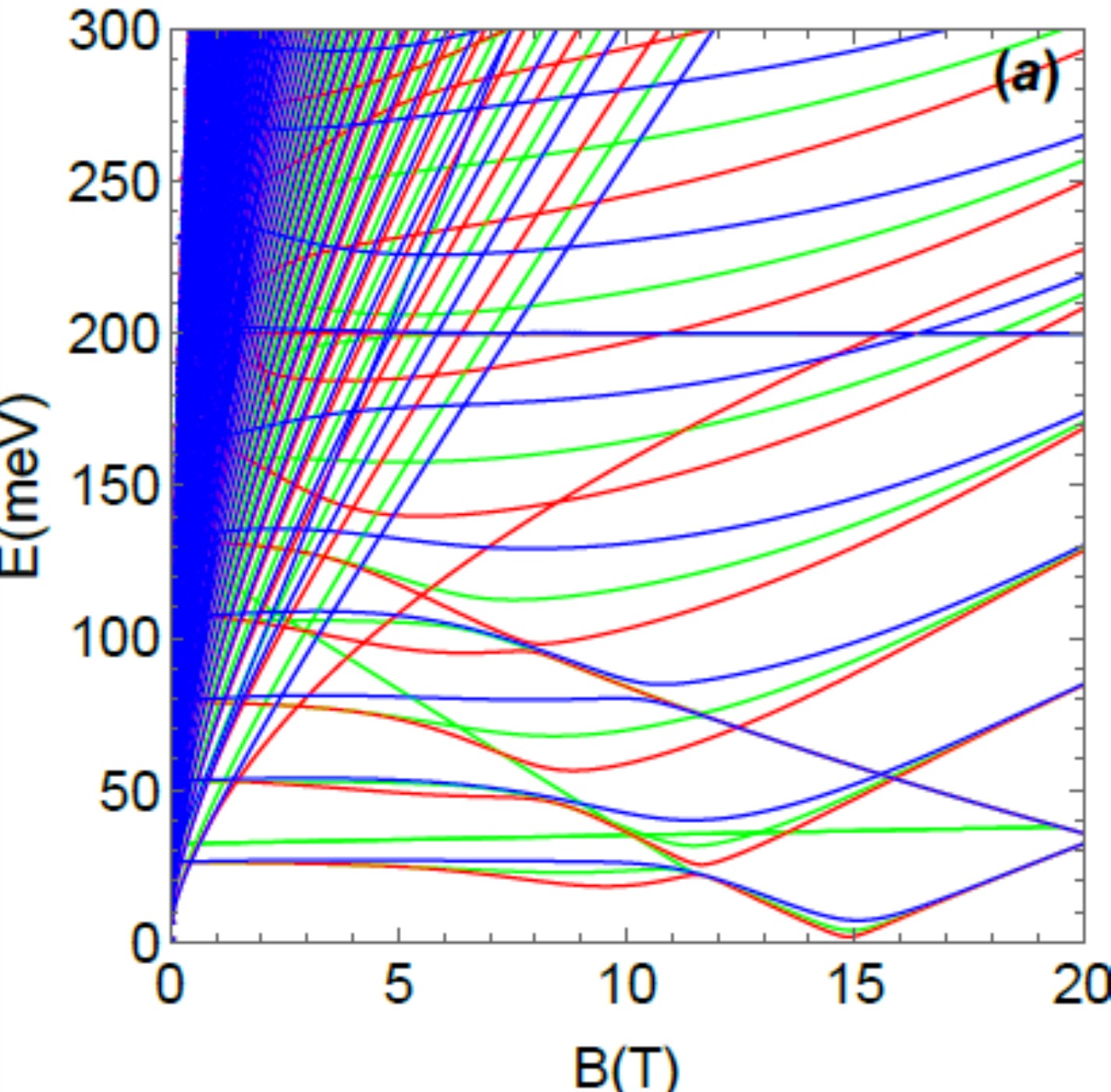}
	\hspace{0.2cm}
	\includegraphics[scale=0.5]{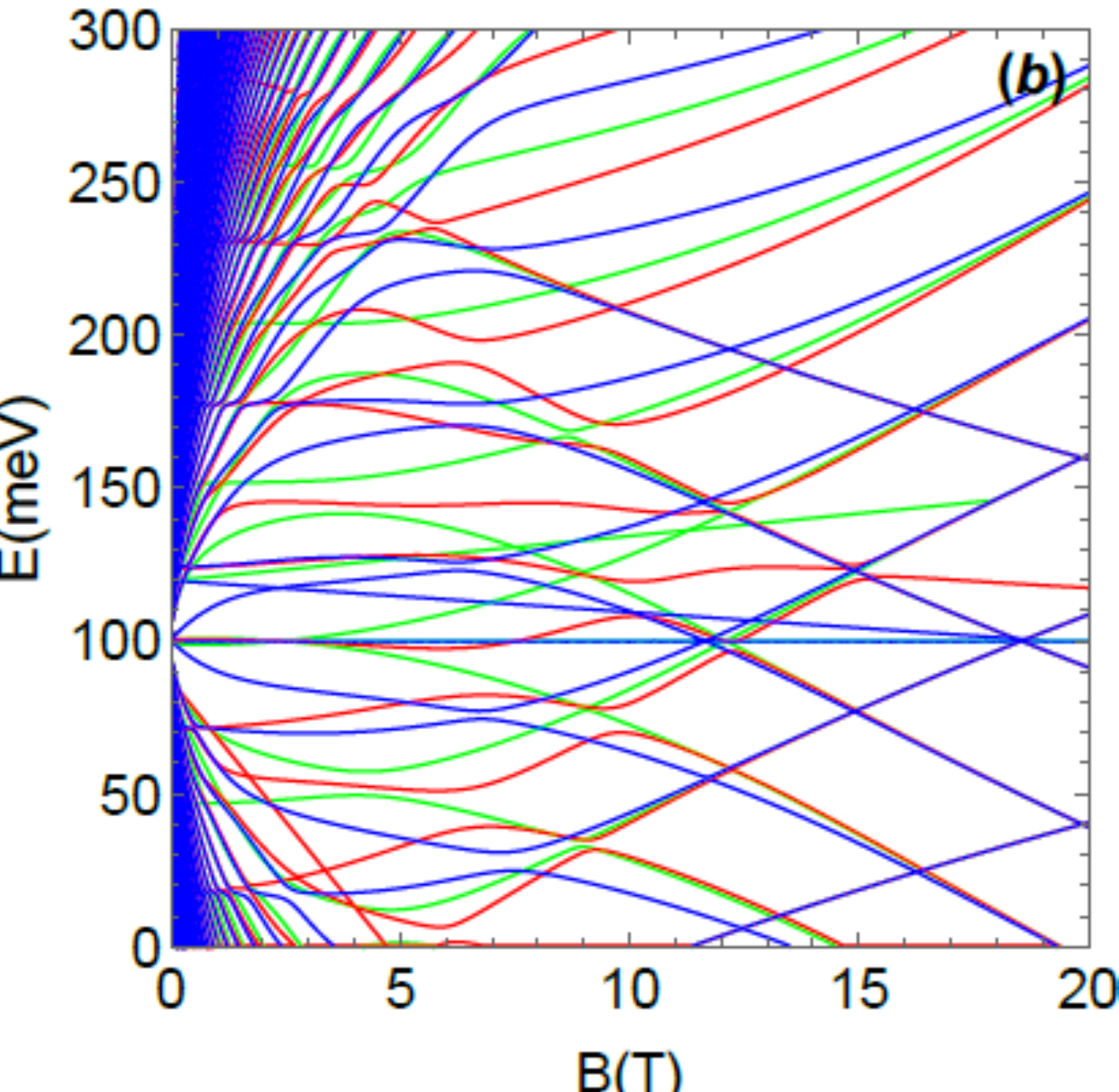}
	\caption{Energy levels of $AA$-stacking bilayer graphene-infinite single layer graphene quantum dot as a function of $B$ with $R = 40$ nm, for the angular momenta $m =0$ (blue), $m=1$ (Red), $m=-1$ (Green), (a): $U_1=U_2= 0$ meV, (b): $U_1=-U_2=100$ meV at the valleys K (solid).}
	\label{f07}
\end{figure}

Figure \ref{f07} shows the energy levels as a function of the magnetic field for (a): $U_1 = -U_2 = 0$ eV, (b): $U_1 = -U_2 = 0.1$ eV and $R=30$ nm, the spectrum is plotted for $m=-1$ (blue), $m=0$(green) and $m=1$ (red). For very small values of $B$, the spectrum becomes degenerate and forms a continuum band. However, the court of increasing magnetic field, the degeneration of the levels is lifted for each angular momentum m and energy levels eventually converge towards the LLs of BLG, This behavior is qualitatively similar to that found in graphene QDs with infinite-mass potential \cite{da Costa14}. The application of a non-zero potential $U_1=U_2=100$ meV (Figure \ref{f07}(b)), shows a significant shift for the energy states compared to the case of the zero potential $U_1=U_2=0$ meV (Figure \ref{f07}(b)).
\section{Conclusion}\label{Conclusion}
We have studied the confinement of charge carriers in two types of systems: SLG-infinite AA-stacked BLG QD and AA-stacked BLG-infinite SLG QD. We have used the continuum model, i.e., the solution of the Dirac-Weyl equation, and obtain analytical results for the energy levels and the corresponding wave functions.

We have implemented the zigzag boundary condition at the SLG-BLG junction to observe the features brought by the zigzag edges in the spectrum. We study the effect of perpendicular electric and magnetic fields on the energy levels.

In the absence of a magnetic field, the energy levels show a  band gap band between the conduction and valence bands.  In the presence of a perpendicular magnetic field, In the case of the SLG-BLG QD interface, when $R\rightarrow 0$ and there is no potential ($U_1=U_2=0$), the specter shows the existence of a non-overlapping band between the valence and conduction bands $ E_k ($m=0$) \neq 0 $, which opens when the point's quantic radius increases.

In contrast to the case of the non-zero potential ($U_1= -U_2\neq 0 $), we have shown a strong dependence of the radius quantum dot, with the appearance of several crossings. We see that the energy spectrum shows two sets of levels, one set for$ R\rightarrow 0$, the energy levels correspond to LLs of bilayer graphene. By increasing R, the second set, for different angular momentum approaches the LL of SLG. In the BLG-SLG QD interface case, with The $R\rightarrow0$ limit, the spectrum corresponds to the LLs of the SLG sheet, being degenerate for angular momenta $m$.  With increasing dot radius, the degeneracy of the levels for different $m$ is lifted, and the levels connect to different LLs of BLG. for certain ranges of dot radius and eventually merge into the LLs of BLG. The spectrum as a function of R for the case biased with $U_1=-U_2=0.1$ eV shows the same behavior as the unbiased case $U_1=U_2=0$ eV with a vertical shift of energy spectrum. We have demonstrated that for very small values  B, the spectrum becomes degenerate and forms a continuum band. However, as the magnetic field increases, the degeneracy of the levels is lifted for each angular momentum.

\end{document}